\documentclass[preprint,prb,showpacs,preprintnumbers,amsmath,amssymb]{revtex4}

\usepackage{graphicx}
\usepackage{dcolumn}
\usepackage{bm}

\begin{document}

\title{Beating of Aharonov-Bohm oscillations
in a closed-loop interferometer} 

\author{Sanghyun Jo$^{1,2}$}
\author{Gyong Luck Khym$^3$}
\author{Dong-In Chang$^{1,2}$}
\author{Yunchul Chung$^4$}
\email{ycchung@pusan.ac.kr}
\author{Hu-Jong Lee$^{1,2}$}
\email{hjlee@postech.ac.kr}
\author{Kicheon Kang$^3$}
\author{Diana Mahalu$^5$}
\author{Vladimir Umansky$^5$}

\affiliation{$^1$Department of Physics, Pohang University of
Science and Technology, Pohang 790-784, Republic of Korea}%
\affiliation{$^2$National Center for Nanomaterials Technology,
Pohang 790-784, Republic of Korea}%
\affiliation{$^3$Department of Physics and Institute for Condensed
Matter Theory,
Chonnam National University, Gwangju 500-757, Republic of Korea}%
\affiliation{$^4$Department of Physics, Pusan National University,
Busan 609-735, Republic of Korea}%
\affiliation{$^5$Department of Condensed Matter Physics,
Weizmann Institute of Science, Rehovot 76100, Israel}%

\date{\today}

\begin{abstract}
One of the points at issue with closed-loop-type interferometers
is beating in the Aharonov-Bohm (AB) oscillations. Recent
observations suggest the possibility that the beating results from
the Berry-phase pickup by the conducting electrons in materials
with the strong spin-orbit interaction (SOI). In this study, we
also observed beats in the AB oscillations in a gate-defined
closed-loop interferometer fabricated on a
GaAs/Al$_{0.3}$Ga$_{0.7}$As two-dimensional electron-gas
heterostructure. Since this heterostructure has very small SOI the
picture of the Berry-phase pickup is ruled out. The observation of
beats in this study, with the controllability of forming a single
transverse sub-band mode in both arms of our gate-defined
interferometer, also rules out the often-claimed multiple
transverse sub-band effect. It is observed that nodes of the beats
with an $h/2e$ period exhibit a parabolic distribution for varying
the side gate. These results are shown to be well interpreted,
without resorting to the SOI effect, by the existence of
two-dimensional multiple longitudinal modes in a single transverse
sub-band. The Fourier spectrum of measured conductance, despite
showing multiple $h/e$ peaks with the magnetic-field dependence
that are very similar to that from strong-SOI materials, can also
be interpreted as the two-dimensional multiple-longitudinal-modes
effect.
\end{abstract}


\maketitle

\section{Introduction}

An electron traversing a path enclosing magnetic flux acquires an
additional phase by the magnetic vector potential. This additional
phase causes the Aharonov-Bohm (AB) oscillation of the electronic
magnetoconductance\cite{AB effect} with the period of $h/e$. The
AB oscillation in solid state devices was first observed by Webb
{\it et al.}\cite{Webb} and Timp {\it et al.}\cite{Timp} in metal
and semiconductor AB rings, respectively. Especially, AB rings
fabricated on the two-dimensional electron gas (2DEG) layer formed
in a semiconductor heterojunction structure like GaAs/AlGaAs have
been studied intensively, because the electron phase coherence in
this system extends much longer than the size of an AB ring.

For instance, a structure of a quantum dot embedded in one arm of
an AB ring was employed recently for accurate measurements of the
phase change of traversing electrons through a quantum dot. Yacoby
\emph{et al.}\cite{Yacoby} first demonstrated that the electron
transport through a quantum dot embedded in a closed-loop AB
interferometer was phase coherent. However, the phase rigidity,
imposed by the Onsager relation in the two-terminal geometry of an
AB ring, hampered measurements of the genuine phase variation
through the quantum dot.\cite{phase rigidity} Later on, Schuster
\emph{et al.}\cite{open-type} measured the phase evolution via a
resonant level of a quantum dot embedded in an open-loop AB
interferometer, which allowed multi-terminal conductance
measurements, while breaking the time-reversal symmetry of the
system. Many other phase measurements through various quantum
structures were followed by using the open-loop AB
interferometer.\cite{phase-measurement} Although open-loop AB
interferometers are more useful for measuring the phase of the
embedded quantum structures, closed-loop AB interferometers are
still in often use because of a simpler theoretical interpretation
of the results.

Detailed studies on closed-loop interferometers, however, often
revealed beats in its AB magnetoconductance oscillation.
\cite{Liu,Bykov} Beats were first explained in terms of mixing of
different multiple transverse sub-band modes arising from a finite
width of a ring.\cite{Liu} The transverse sub-band modes, with the
conductance of $2e^2/h$ per mode, are the quantum states defined
by the transverse potential in the interferometer, which can be
tuned by the voltages applied to the gates, {\it i.e.}, quantum
point contacts. This model, however, failed to explain the
appearance of beats in the conductance with a single transverse
sub-band (STSB) mode, which was observed by the same
group.\cite{Liu} Later, Tan {\it et al.}\cite{Tan} pointed out
that the clockwise and the counterclockwise moving electron states
in an STSB mode may generate different AB-oscillation frequencies,
producing the beats reported in Refs. 8 and 10. In the meantime,
similar beating effect was observed in the AB interferometry using
two-dimensional electron gas (2DEG) [two-dimensional hole gas
(2DHG)] systems based on strong spin-orbit interaction (SOI)
materials such as InAs and InGaAs [GaAs]
heterojunctions.\cite{SO1,SO2,SO3,SO4} It was interpreted as the
evidence for the revelation of the Berry phase \cite{Berry} caused
by the strong SOI. In this case, the Fourier spectrum of the beat
pattern showed multiple splitting of the $h/e$ peak. Morpurgo {\it
et al.}\cite{SO1} and Meijer {\it et al.}\cite{SO2} showed clear
splitting of the $h/e$ peak into multiple peaks in the
ensemble-averaged Fourier spectrum over several measurements for a
fixed and varied gate voltage(s), respectively. They attributed
the splitting to the Berry-phase \cite{Berry} pickup by the
traversing electrons due to the SOI. Yau {\it et al.}\cite{SO3}
and Yang {\it et al.}\cite{SO4} also observed the multiple
splitting of a single Fourier-spectrum peak and the evolution of
the spectrum for varying magnetic fields. Based on the numerical
simulation, the authors also attributed them to the result of the
Berry-phase pickup. All these works, however, did not consider the
effect of possible mixing of multiple conducting channels.

In this study, we investigated the behavior of the
magnetoconductance from a closed-loop AB interferometer fabricated
on a GaAs/Al$_{0.3}$Ga$_{0.7}$As heterostructure 2DEG, with a very
weak SOI. The magnetoconductance always revealed clear beats of AB
oscillations. The corresponding Fourier spectrum exhibited
multiple peaks closely spaced around the field value where the
$h/e$ peak was expected. The patterns were very similar to those
observed previously from the AB rings fabricated on strong SOI
materials. To interpret our observation of the beats in our system
with a very weak SOI, we simulated two-dimensional (2D) AB ring
using the method adopted by Tan \emph{et al.}\cite{Tan} In the
simulation, even in a fixed transverse sub-band mode, additional
eigen-modes existed to contribute to the electron transport as a
function of the Fermi level and a magnetic field. For the
remainder of this paper these modes will be referred as
longitudinal modes. These modes were found to evolve in different
ways depending on the geometry and thus induce beats of AB
oscillations in a 2D interferometer. We were able to explain all
our results in terms of the formation of the 2D multiple
longitudinal modes in an STSB without resorting to the strong-SOI
effect. In fact, measurements of the weak localization (WL) and
the Shubnikov-de Haas (SdH) oscillation in our system revealed
that the SOI was almost negligible. In our observation the $h/2e$
periodicity was noticeable only around the nodes of the beats. In
addition, the nodes of the beats formed a parabolic distribution
for varied voltages applied to one of the loop-forming side gates,
which was not observed in the previous
studies.\cite{Liu,Tan,Bykov} These two general features of our
data were in accordance with the prediction of the 2D
multiple-longitudinal-modes effect in an STSB.

\section{Results and Discussion}

\subsection{Device structure and measurements}

Figure 1 shows a gate-defined closed-loop AB interferometer,
fabricated on a wafer of GaAs/Al$_{0.3}$Ga$_{0.7}$As
heterostructure 2DEG. The electron density of the 2DEG was about
$2\times10^{11}$ cm$^{-2}$ with the mobilities higher than $2
\times 10^6$ cm$^2$ V$^{-1}$s$^{-1}$ at 4.2 K. The 2DEG layer
resided 70 nm below the surface of the wafer. The interferometer
was defined by an island gate, four side gates, and two
quantum-dot gates as shown in the figure. The shape of this
structure is similar to the one reported previously.\cite{Yacoby}
The electron path inside the interferometer was defined by tuning
negative voltages to these gates. Unlike a closed-loop
interferometer defined by the mesa formation,
\cite{Liu,Bykov,SO1,SO2,SO3,SO4} the number of transverse
conducting channels (sub-bands) in each (left or right) arm could
be fine-tuned by the six side gates. This was a big advantage over
mesa-defined AB rings, where the existence of multiple transverse
sub-bands was inevitable in practice. In our study, the number of
transverse sub-bands for each path was kept close to or below one
to avoid any possible interference among electrons from different
transverse sub-bands. The quantum dot embedded in the
interferometer in Fig. 1 was not formed during measurements of
beats, despite the physical presence of the gates to form it in
the sample. All the measurements were made at the base temperature
of 10 mK in a dilution fridge. The corresponding electron
temperature estimated from measurements of the Coulomb blockade in
the quantum dot \cite{electron temp} was around 140 mK. The
standard phase-sensitive lock-in technique was employed to improve
the signal to noise ratio. The AB magnetoconductance oscillations
were taken by observing the current between the two ends (the
source and the drain) of the AB ring in an excitation voltage of
10 $\mu$V rms. For measurements of the WL and the SdH oscillation,
a four-probe configuration was employed, where a 15-nA rms current
was applied between the source and the drain while monitoring the
corresponding voltage difference.

\subsection{Gate-voltage dependence of beats}

The magnetoconductance of the interferometer at various voltages
of Side-gate 1 ($V_{sg1}$) was measured by applying a
10-$\mu$V-rms excitation voltage to the interferometer, while
monitoring the source-drain current. The magnetic field was swept
in the range between -1500 G and 1500 G, while the $V_{sg1}$ was
varied from -271.35 mV to -291.15 mV at intervals of 1.65 mV.
Since the conductance for any $V_{sg1}$ was lower than $2e^2/h$ it
was assumed that less than an STSB \cite{STSB} was open for each
path of the interferometer. Since the visibility of the AB
oscillation was about 2\% the background current level was
subtracted for clarity. As shown in Fig. 2(a) the field dependence
of the conductance exhibits clear beats of AB oscillations for any
values of $V_{sg1}$. In addition, the phase evolved
discontinuously in the variation range of $V_{sg1}$. This
phenomenon of the rigid phase has been observed by
others\cite{Yacoby,closed-type} and interpreted by the Onsager
relation in a two-terminal closed-loop interferometer.\cite{phase
rigidity} The relation stipulates that the electronic conductance
through a closed-loop interferometer should be symmetric with
respect to the $B$=0 axis, allowing only 0 or $\pi$ phase. As a
result, the phase change occurs only by $\pi$ via a
double-frequency ($h/2e$ period) AB oscillation
regime.\cite{Yacoby,phase rigidity} The expected phase change in a
one-dimensional (1D) closed-loop interferometer, with an STSB per
path without considering finite width of the electron path, is
shown in Fig. 2(b). As in the figure, the phase stays unaltered
for the variation of the Fermi energy ($E_F$) in the leads until
it suddenly changes by $\pi$. The simulation also suggests that
the frequency of the AB oscillation doubles with $h/2e$
periodicity between the transition of the two phase states. For a
1D closed-loop interferometer, with an STSB in each arm, this
transition should occur at the same $E_F$ in all magnetic fields.
In other words, as in Fig. 2(b), the transition region should be
in parallel with the $B$-field axis. In our measurements, however,
the transition region distributes rather parabolically in the
$V_{sg1}$-vs-$B$-field plot [Fig. 2(c)]. These features were
reproducible even after thermal recycling. Refs. 10 and 14 report
the existence of $h/2e$ period in the beats of AB oscillations,
but without this parabolic distribution. Similar behavior was
observed with the variation of the source-drain bias voltage. A
parabolic distribution of the transition regions was also observed
when the voltage of the island gate was varied, while fixing the
voltages for all the other gates used to form the interferometer.

\subsection{Strength of the spin-orbit interaction in GaAs/Al$_{0.3}$Ga$_{0.7}$As heterostructure}

Such a beating effect without the parabolic distribution of the
$h/2e$ transition region mentioned above has been observed in
various systems like metallic AB rings,\cite{Webb,metal-ring}
GaAS/AlGaAs 2DEG AB rings,\cite{Liu,Bykov} GaAs/AlGaAs 2DHG
rings,\cite{SO3} InAs-based\cite{SO1,SO4} and
InGaAs-based\cite{SO2} 2DEG rings. Although the beating features
are similar to each other, interpretations of the origin of the
beats are varied. Two main models suggested to explain the beats
are based on picking up the Berry phase induced by the SOI and
mixing of multiple conducting channels caused by the finite width
of an AB ring involved. The Berry-phase-related interpretation of
the beats is favored for the AB rings formed on the relatively
strong SOI systems like GaAs/AlGaAs-based 2DHG, and InAs-based and
InGaAs-based 2DEG's. In contrast, the model of mixed multiple
conducting channels is favored for AB rings in metallic films and
GaAs/AlGaAs-based 2DEG's with relatively small
SOI.\cite{Small-SOI}

To estimate the strength of the SOI in our system, both the WL and
the SdH oscillation were measured on an identical sample that was
used to measure the beats. Fitting the anti-localization dip in
the magnetoresistance for strong SOI materials leads to resolving
the SOI strength.\cite{WL} By contrast, the SOI gives two
different frequencies for the SdH oscillation. It results in beats
in the SdH oscillation, allowing one to estimate the SOI strength.
\cite{SdH} Figs. 3(a) and 3(b) illustrate the magnetoresistances
of the WL and the SdH oscillation, respectively, measured in a
Hall-bar geometry formed on our sample. All the gates in the
interferometer were grounded during the measurements. Fig. 3(a)
shows the magnetoresistance averaged over 37 measurements with the
field resolution of 0.2 G, which exhibits no discernible
anti-localization dips. Fig. 3(b) shows the measured SdH
oscillation, without any beats, either. These results strongly
imply that the SOI was very weak or almost negligible in the
GaAs/Al$_{0.3}$Ga$_{0.7}$As heterostructure 2DEG of our
interferometer. Although it is known that metallic gates may
enhance the strength of the SOI, the enhancement is usually less
than 30$\%$ of the original value.\cite{gatecontrol-SOI} We thus
expect that the SOI caused little effect on our magnetoconductance
measurements.

\subsection{Two-dimensional multiple-longitudinal-modes effect}

In our measurements, only an STSB was set to open in each arm. The
beats of AB oscillation for the STSB conduction were
experimentally observed previously,\cite{Liu,Bykov} where the
beats were theoretically interpreted in terms of mixing of
clockwise and counterclockwise moving states of electrons with
different frequencies \cite{Tan} in an STSB mode. We employed the
model as used in Ref. 9 to simulate the beats and the evolution of
their transition region as a function of $V_{sg1}$. The result
indicates that the multiple longitudinal modes existing in an STSB
mode generate the AB oscillations. However, the multiple
longitudinal modes give more than two oscillation frequencies,
with the number of the oscillation frequencies varying with the
magnetic field, due to the finite width of the electron path
inside the interferometer. The magnetic field dependence will be
discussed in the next section in more detail.

We adopted the same form of Shr\"{o}dinger equation as used in
Ref. 9, but with varied parameter values to fit our experimental
conditions. The eigenvalues are

\begin{eqnarray}
E_{n,m}(B)&=&(n+\frac{1}{2}+\frac{1}{2}\sqrt{m^2+\frac{{\eta}^2}{4}})\hbar\omega
+ \frac{m}{2}\hbar\omega_c - \frac{\eta}{4}\hbar\omega_0,\nonumber\\
n&=&0,1,2,3,\cdot\cdot\cdot, \ \ m \ = \
\cdot\cdot\cdot,-1,0,1,\cdot\cdot\cdot,
\end{eqnarray}

\noindent where $\omega_c=eB/\mu$,
$\omega=\sqrt{{\omega_c}^2+{\omega_0}^2}$,
$\eta=\mu\omega_0{r_0}^2/\hbar$, $\mu$ is the effective mass of
the electron, $\omega_0$ is the transverse confinement strength of
the gates defining the electron path, and $r_0$ is the average
radius of the interferometer. Here, the quantum number $n$ is the
transverse sub-band index and $m$ is related with the longitudinal
motion.\cite{Tan} In our measurements, we tuned the interferometer
to have an STSB, corresponding to $n$=0. In Fig. 4(a), the energy
band $E_{n,m}(B)$ is shown for an STSB mode with $n=0$ and
$|m|\leq$30.

As shown in Fig. 4(a), several conducting modes exist even for an
STSB mode because of the existence of multiple longitudinal modes.
The corresponding energy eigenvalues depend on $m$ as well as $n$.
For this dependence, the dispersion of the bands is asymmetric
about the band minima. The minima shift to higher energies for
nonzero magnetic fields,\cite{Tan} which result in breaking of the
discrete symmetry under the translation along $B$-axis. As
indicated by filled (open) circles in the figure, this makes band
crossing (splitting) points respond nonlinearly to the applied
magnetic field. Fig. 4(b) is an expanded band diagram of the
region in the dashed  box in Fig. 4(a). The band-splitting points
show a clear downturn curvature as represented by open circles,
the shape of which is very similar to that of the conductance with
the $h/2e$ period in Fig. 2(c).

The conductance for a given Fermi energy $E_F$ in the dashed-box
region in Fig. 4(a) was calculated using the Landauer-B\"{u}ttiker
formula\cite{Landauer-Buttiker}

\begin{eqnarray}
G(E_F, B)&=&\frac{2e^2}{h}\sum_{n,m}\frac{{\Gamma}^2}{[E_F -
E_{n,m}(B)]^2 + {\Gamma}^2},
\end{eqnarray}

\noindent where only $n=0$ is taken into account and the
transmission probability via an eigen mode is given by the
Breit-Wigner formula at zero temperature. Here, $\Gamma$ is the
coupling constant between the emitter (or the collector) and the
AB ring. The $\Gamma$ was set at 3\% of $\hbar\omega_0$, which
corresponds to a weak coupling regime of the system. The
calculated results are shown in Fig. 5(a). The conductance shows a
clear beat pattern with a variation of the envelope as a function
of $E_F$. Fig. 5(b) shows similar behavior observed experimentally
for varied $V_{sg1}$. The figures indicate that the $h/2e$ ($h/e$)
oscillation predominates in the nodes (antinodes) of the beats.
The region of dominant $h/2e$ oscillation appears in different
magnetic fields as $E_F$ is varied. A grey-scale plot of the
calculated conductance is shown in Fig 6. The $E_F$ was set to the
range in the dashed square box in Fig. 4(a). The band shown in
Fig. 4(b) is overlapped on the grey-scale plot by white solid
lines. Contrary to the results calculated for an STSB in a 1D
interferometer as in Fig. 2(b), the region where the nodes of the
beats appear shows an $h/2e$ oscillation and distributes
parabolically in an $E_F$-vs-$B$-field plot. This result is in
qualitative agreement with our observation shown in Fig. 2(c).
Clearer revelation of the double frequency region in the
simulation may be from smaller $\Gamma$ value and higher
visibility of AB oscillations than those in the measurement. The
parabolic evolution of the node region can be naively understood
with the band diagram shown in Fig. 4. According to Eq. (2) the
overall conductance through an AB ring is determined mainly by the
bands located close to $E_F$. Hence, if $E_F$ is at a crossing
point of two bands (marked by a filled circle in the figure) the
$h/e$ oscillation becomes predominant. By contrast, the $h/2e$
oscillation predominates if $E_F$ is at a band splitting point
marked by an open circle in the figure. The conductance itself is
bigger for $h/e$ oscillation, since two longitudinal modes
simultaneously contribute to the conductance, while a single
longitudinal mode contributes to the conductance for $h/2e$
oscillation. This is the reason why $h/e$ ($h/2e$) oscillation is
dominant in the antinodes (nodes). Since these band crossing
(splitting) points evolve parabolically as a function of $E_F$ and
$B$, the antinodes (nodes) also evolve in the same fashion. It
should be emphasized that the finite width of the electron path in
an AB interferometer is responsible for the parabolic evolution of
the band crossing points. According to our simulation, however, a
parabolic evolution with an upturn curvature is also possible in
different regions of $E_F$ (not shown). Similar experimental
result was also observed in a different region of the gate
voltages.

The beat feature similar to Fig. 5 was reported in mesa-defined
closed-loop AB interferometers,\cite{Liu,Bykov} with the results
interpreted in terms of 2D multiple-longitudinal-modes
effect\cite{Tan} in an STSB. In these works, however, no parabolic
evolution of the beat pattern was reported. The parabolic
evolution gives an additional confirmation of the validity of the
model and better understanding of the 2D
multiple-longitudinal-modes effect in an STSB. Observing the
smooth evolution of the node region was possible in our study,
because the number of transverse sub-bands was fine-controlled by
the six individual gates, rather than by a single gate covering
the whole mesa-defined AB ring as used in the other measurements.
We tuned all the gates individually to leave an STSB open. Only
one gate was then slightly modified to observe such an evolution.

\subsection{Fourier transform analysis and comparison with results from an AB ring of strong SOI}

The Fourier transform analysis is a very effective way to analyze
the beating of AB oscillation induced both by the
multiple-conducting-channel effect\cite{Liu,Bykov} and the SOI
effect.\cite{SO1,SO2,SO3,SO4} The Fourier spectrum gives the
information on different conducting channels (or modes) inside an
interferometer with very small frequency differences. In this
section, we present the Fourier spectrum of two representative
beats, which show the antinodes and the nodes at $B$ = 0. The
zero-padding technique\cite{zero-padding}, that was used to
analyze the beats induced by the SOI effect,\cite{SO3,SO4} was
adopted for our Fourier transform analysis. The magnetoconductance
data taken in the range between -1500 G and 1500 G were used for
the analysis, because SdH oscillation started to appear at around
$\pm$ 2000 G due to the high mobility of electrons in the sample.
Within this range, four different widths of the Hamming
window\cite{Hamming-Window} ($\pm$500 G, $\pm$900 G, $\pm$1200 G
and $\pm$1500 G) were used. This allowed a close examination of
any field-dependent effect.

The left (right) panel of Fig. 7(a) is the Fourier spectrum for
the antinode (node) at $B$=0. These two cases show different
features, but their magnetic field dependencies are similar to
each other. When the antinode is at $B$=0, the Fourier spectrum
shows a main center peak and a trace of small side peaks in a low
magnetic field range. With increasing magnetic field, the main
center peak starts to split, while the side peaks become clearer.
When the node is at $B$=0, however, the amplitudes of the main
center peak and symmetric side peaks are comparable even in a
low-magnetic-field range. With increasing magnetic field in this
case, the center peak starts to split, while the amplitudes of the
side peaks increase. Fig. 7(b) is the Fourier spectrum of the
calculated magnetoconductance when an antinode and a node are at
$B$=0. The features are similar to those of the experimental
Fourier spectrum.

The evolution of the Fourier spectrum as a function of the
magnetic field can be naively explained as follows. First, the
band crossing points marked by filled circles in the figure 4(a)
give a Fourier-transform peak (the center peak). Second, open
circles marked in the figure give two Fourier-transform peaks (the
side peaks), where one is slightly lower and another is slightly
higher than the peak given by the band crossing points. These are
major three peaks shown in the spectrum. The frequency difference
and the field dependence of the peaks are caused by the breaking
of the discrete symmetry under the translation along the $B$ axis.
Even with many energy bands around $E_F$, the conductance is
mainly determined by the bands close to $E_F$. Thus, if the data
showing an antinode at $B=0$ are taken for the Fourier spectrum
analysis with a very narrow magnetic-field window only a center
peak appears in the spectrum. As the width of the magnetic field
window increases, the nodal part is included in the analysis to
produce the side peaks in the spectrum. By the same analogy, if
the data showing node at $B=0$ are taken for the analysis with a
narrow magnetic field window only the side peaks appear in the
spectrum. As the antinodal part is included in the analysis by
enlarging the magnetic field window the center peak appears in the
spectrum. These features are illustrated in Fig. 7(a). However,
the experimental data given in the right panel of Fig. 7(a) show
three peaks (a center peak and two side peaks) even in a
low-magnetic-field range. That is because the magnetic field
window is wide enough to contain the antinodes as well as the node
as shown in Fig. 2(a).

Similar evolution of Fourier-spectrum peaks as a function of
magnetic field was also reported in Refs. 13 and 14. Those were
interpreted as evidences for the Berry phase pickup by the SOI,
although the interpretation in Ref. 13 has been
controversial.\cite{SO3-Reply1,SO3-Reply2} We believe that the
evolution of peaks observed in the Fourier spectrum as a function
of magnetic fields are not a conclusive evidence of the spin-orbit
coupling,\cite{SO4} since the same effect is also observable in
systems with negligible spin-orbit coupling. The beats in this
case can be explained by the 2D effects inside the interferometer
as clearly demonstrated in our study, both in experiment and in
simulation. Meanwhile, the center-peak splitting and the side
peaks were also observed in other magnetoconductance measurements
using AB rings fabricated on strong SOI materials.\cite{SO1,SO2}
In these works, the splitting of the center peak was focused and
regarded as an evidence for the Berry-phase pickup by the SOI. It
was pointed out, however, that the splitting was too large to be
explained by the Berry phase pickup only. In the works, the
existence of the side peaks was not clearly explained, either. As
shown in our data, the splitting of the center peaks in Fig. 7(a)
is around 0.001 G$^{-1}$, which is comparable to those in Refs. 11
and 12. Even with almost negligible SOI in our system, the
features of the Fourier spectrum of the measured
magnetoconductance show a similarity to those found in strong-SOI
systems.\cite{SO1,SO2,SO3,SO4} Our study demonstrates that the
features in the Fourier spectrum can be generated irrespective of
the Berry-phase pickup.

\section{Summary and Conclusion}
The beats of AB oscillations in a closed-loop interferometer
fabricated on GaAs/Al$_{0.3}$Ga$_{0.7}$As 2DEG were observed.
Unlike mesa-defined interferometers, in our measurements, the
electron path in each arm of the interferometer was kept less than
a single transverse sub-band (STSB) by the fine control of six
individual path-forming side gates. This enabled us to study the
effect of the two-dimensional longitudinal modes developing in an
STSB induced by the external magnetic field. Our simulation
indicates that the experimentally observed beats of AB
oscillations resulted from the multiple oscillation frequencies
induced by finite width of the interferometer even with an STSB
mode. The $h/2e$ AB oscillations, imposed by the Onsager relation
in our closed-loop two terminal geometry, appeared in the nodes of
beats, while usual $h/e$ AB oscillations appeared in the
antinodes, which was in agreement with our simulation results. It
was also found that nodes of the beats distributed parabolically
with varying the voltages of the side gates. We confirm that the
phenomena can be explained by the 2D multiple-longitudinal-modes
effect in the presence of an STSB, even without resorting to the
strong spin-orbit interaction.

\section*{ACKNOWLEDGMENTS}
This work was supported by Electron Spin Science Center in Pohang
University of Science and Technology (POSTECH) and Pure Basic
Research Grant R01-2006-000-11248-0 administered by KOSEF. This
work was also supported by the Korea Research Foundation Grants,
KRF-2005-070-C00055 and KRF-2006-312-C00543, and by POSTECH Core
Research Program.

\newpage
\textbf{FIGURE CAPTIONS}
\\
\\
Figure 1. SEM image of the gate-defined closed-loop Aharonov-Bohm
interferometer fabricated on a GaAs/Al$_{0.3}$Ga$_{0.7}$As
heterostructure two-dimensional electron-gas layer.
\\
\\
\\
Figure 2. (color) (a) The magnetoconductances for various voltage
to Side-gate 1, $V_{sg1}$, after subtracting the smooth
backgrounds from the raw data, as a function of magnetic fields.
The magnetic field was varied from -500 G to 500 G, while
$V_{sg1}$ from -271.35 mV (top) to -291.15 mV (bottom) at
intervals of 1.65 mV. (b) The phase rigidity expected in a
one-dimensional closed-loop AB-ring is illustrated. (c) The
two-dimensional pseudocolor plot of the magnetoconductance shown
in (a). A double frequency regime is guided by the dashed line.
For both pseudocolor plots, the red (blue) color represents the
high (low) magnetoconductance.
\\
\\
\\
Figure 3. (a) Normalized magnetoresistance averaged over 37
independent measurements and (b) the longitudinal Hall resistance
of a Hall-bar structure, both plotted as a function of magnetic
fields.
\\
\\
\\
Figure 4. (color online) (a) The calculated energy-band diagram of
a two-dimensional AB ring. The energy eigenvalues for $n=$0 and
${|m|\leq}$30 modes are plotted as a function of magnetic fields.
Multiple bands correspond to different $m$ values. The filled
circles denote the band-crossing points, while the open circles
denote the middle point between two neighboring band-crossing
points. (b) The band diagram of the dashed-box region in (a),
enlarged for clarity.
\\
\\
\\
Figure 5. (a) The magnetoconductance calculated for $E_F$=0.306
meV and 0.298 meV, where the Fermi energies were set to lie in the
lowest sub-band to simulate the measurement condition. (b)
Measured magnetoconductance for $V_{sg1}$=-274.65 mV and -287.85
mV.
\\
\\
\\
Figure 6. (color online) Grey-scale plot of the conductance,
calculated for a two-dimensional electron-gas AB-ring, as a
function of $E_F$ and magnetic field, where the band shown in Fig.
4(b) is overlapped in white solid lines. The white dashed line is
a guide to denote the region where the $h/2e$ oscillation is
dominant.
\\
\\
\\
Figure 7. (color online) (a) The Fourier spectrum of the $h/e$
magnetoconductance measured at different $V_{sg1}$ in a few
magnetic-field windows. In the left (right) panel, with $V_{sg1}$
=-289.50 mV (-277.95 mV), the antinode (node) of beats was at
$B$=0. In the left (right) panel, the magnetic field was varied in
the ranges of $\pm$500 G ($\pm$500 G), $\pm$900 G ($\pm$1200 G),
and $\pm$1200 G ($\pm$1500 G), indicated by the thick solid line,
the thin solid line, and the dotted line, respectively. (b) The
Fourier spectrum of $h/e$ magnetoconductance calculated for
different values of $E_F$ and a few magnetic-field windows are
shown for comparison. In the left (right) panel, with $E_F$=0.297
meV (0.305 meV), the beating features for $B$=0 are similar to
those of (a). In the left (right) panel, the magnetic field was
varied in the ranges of $\pm$800 G ($\pm$800 G), $\pm$1200 G
($\pm$1600 G), and $\pm$1600 G ($\pm$2400 G), indicated by the
thick solid line, the thin solid line, and the dotted line,
respectively.

\newpage

\begin{figure}[t]
\begin{center}
\leavevmode
\includegraphics[width=.85\linewidth]{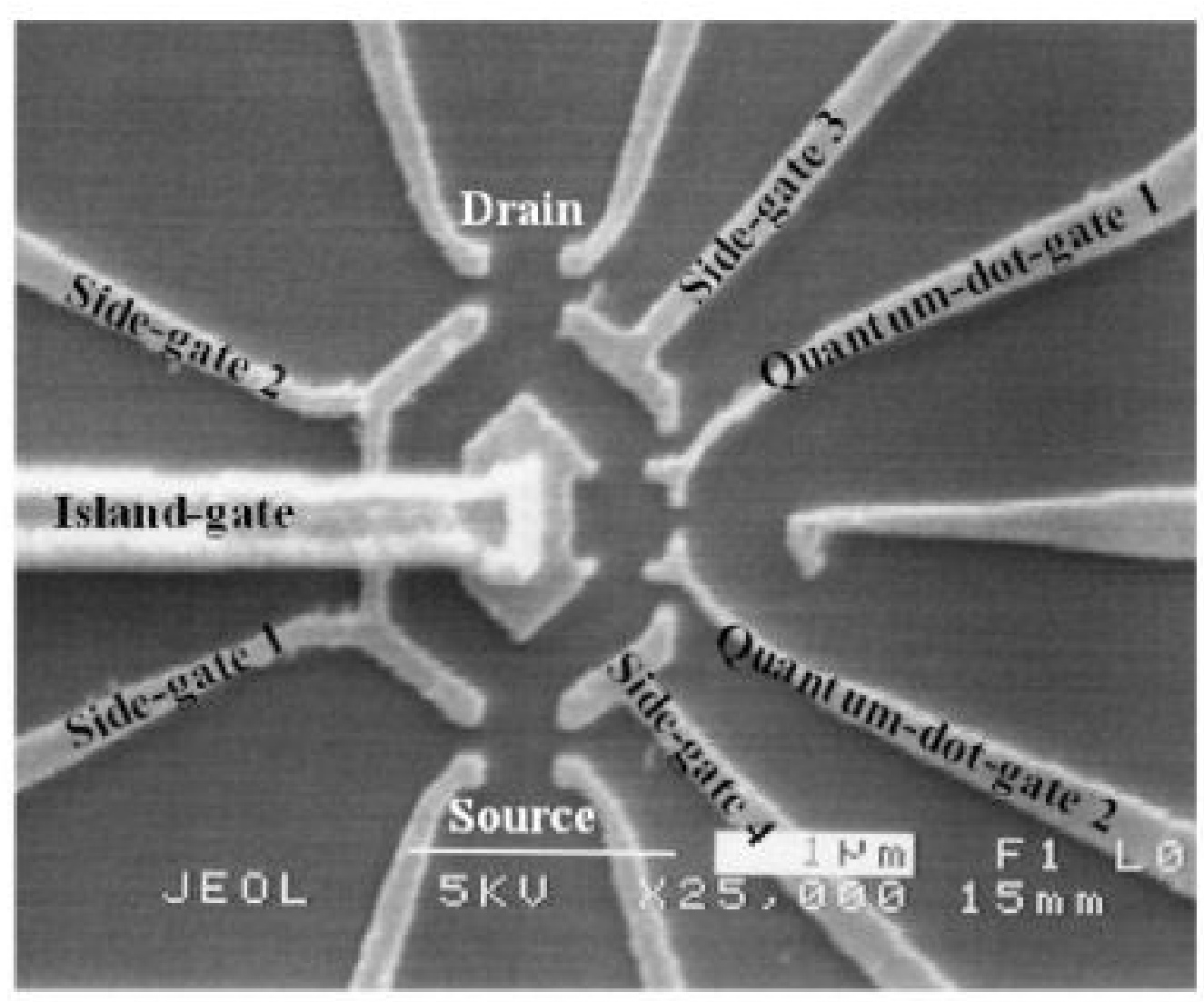}
\caption{}
\end{center}
\end{figure}

\begin{figure}[t]
\begin{center}
\leavevmode
\includegraphics[width=0.6\linewidth]{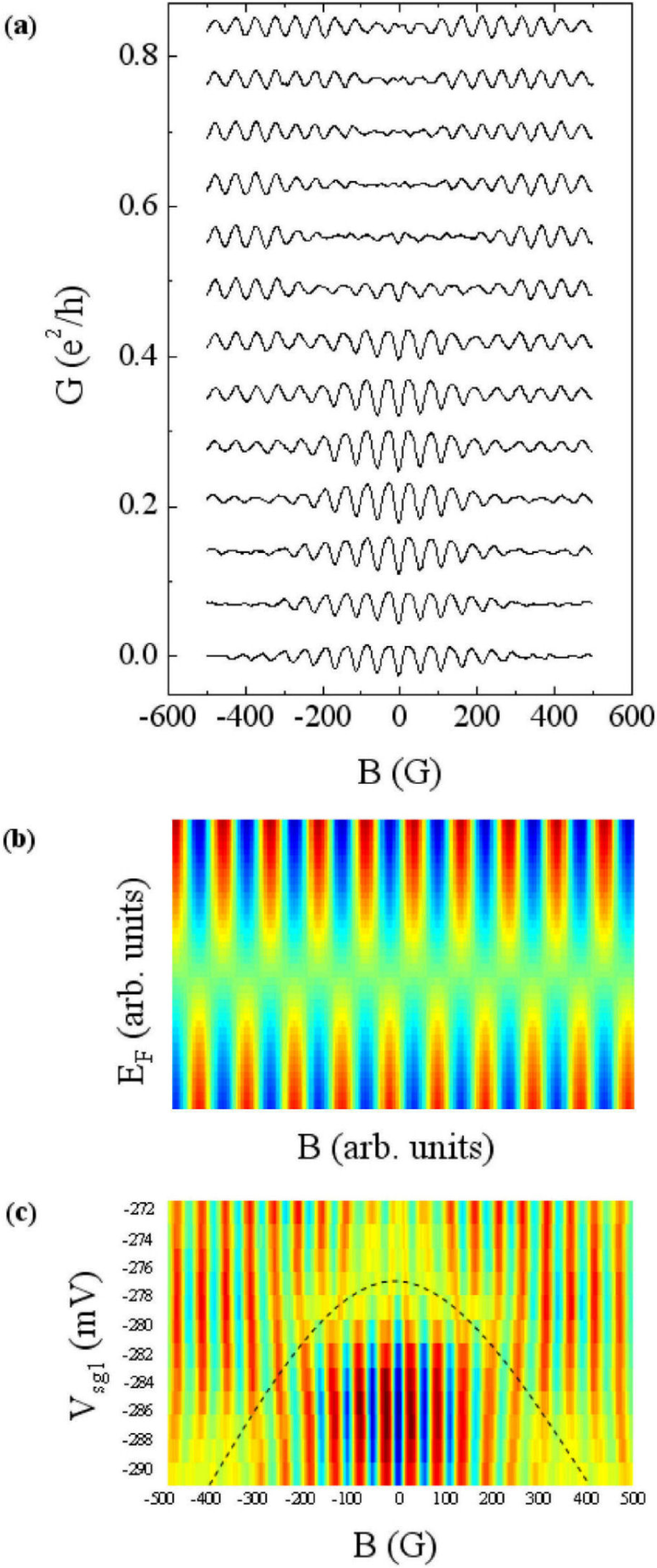}
\caption{}
\end{center}
\end{figure}

\begin{figure}[t]
\begin{center}
\leavevmode
\includegraphics[width=.7\linewidth]{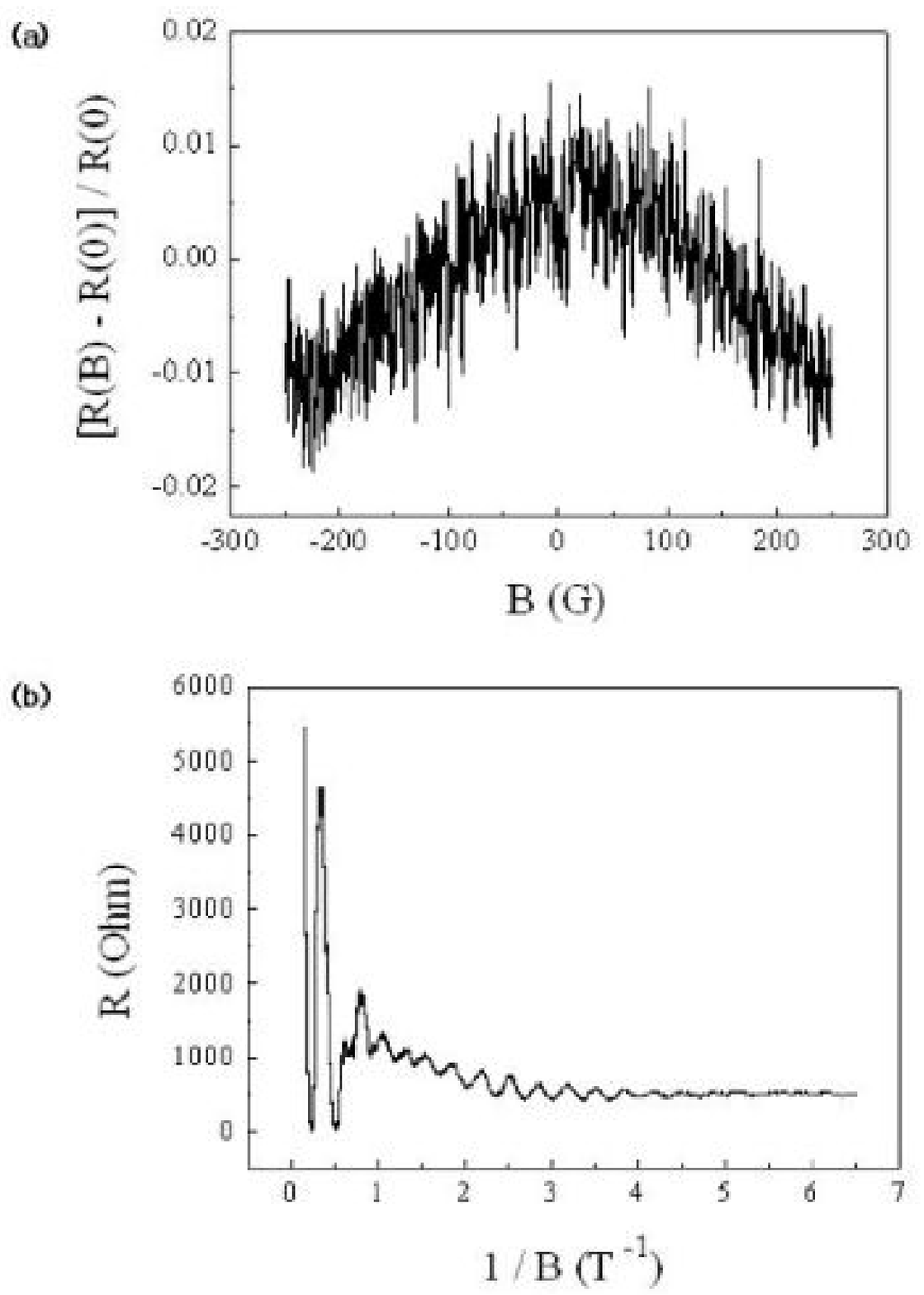}
\caption{}
\end{center}
\end{figure}

\begin{figure}[t]
\begin{center}
\leavevmode
\includegraphics[width=0.8\linewidth]{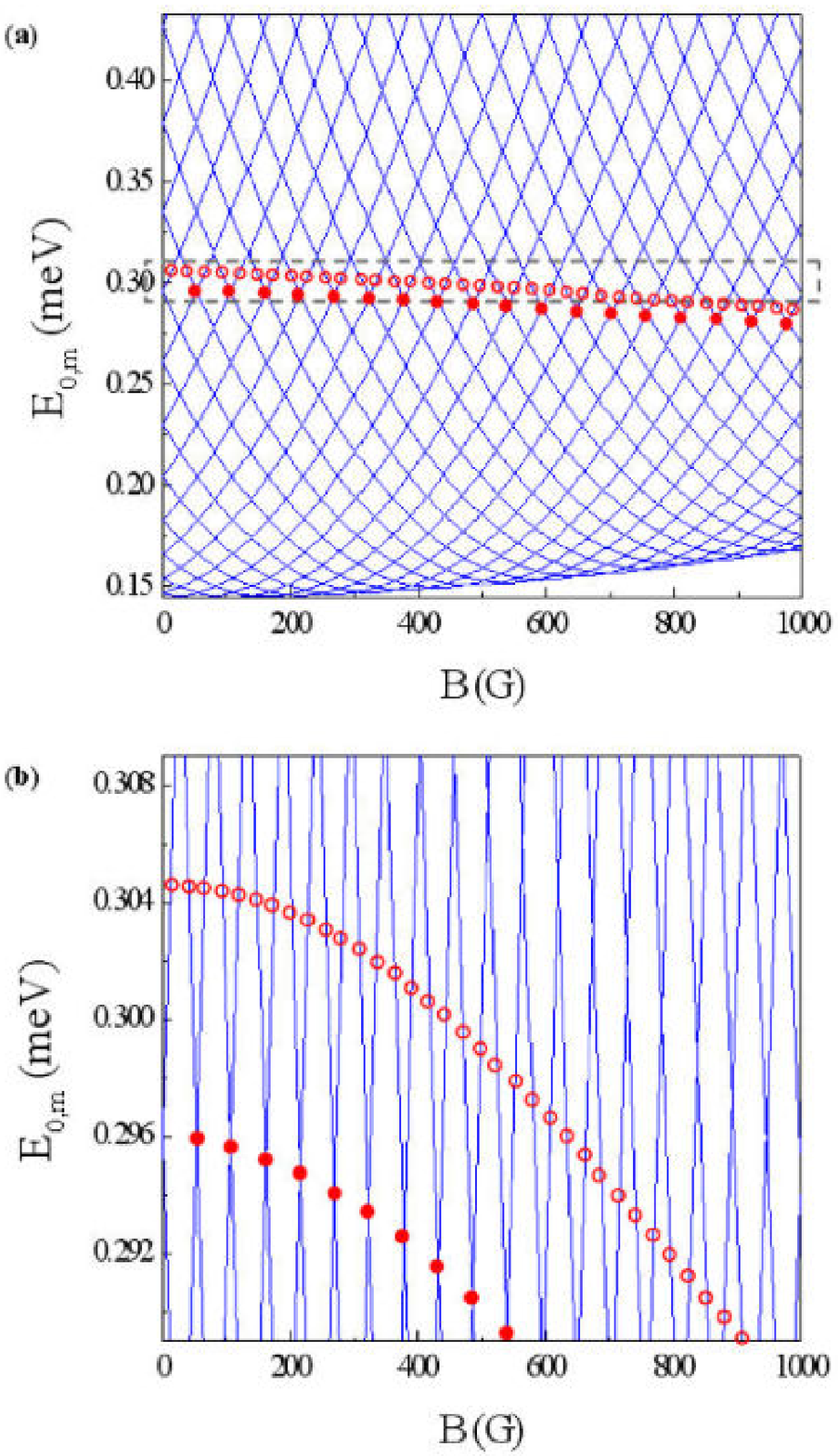}
\caption{}
\end{center}
\end{figure}

\begin{figure}[t]
\begin{center}
\leavevmode
\includegraphics[width=1\linewidth]{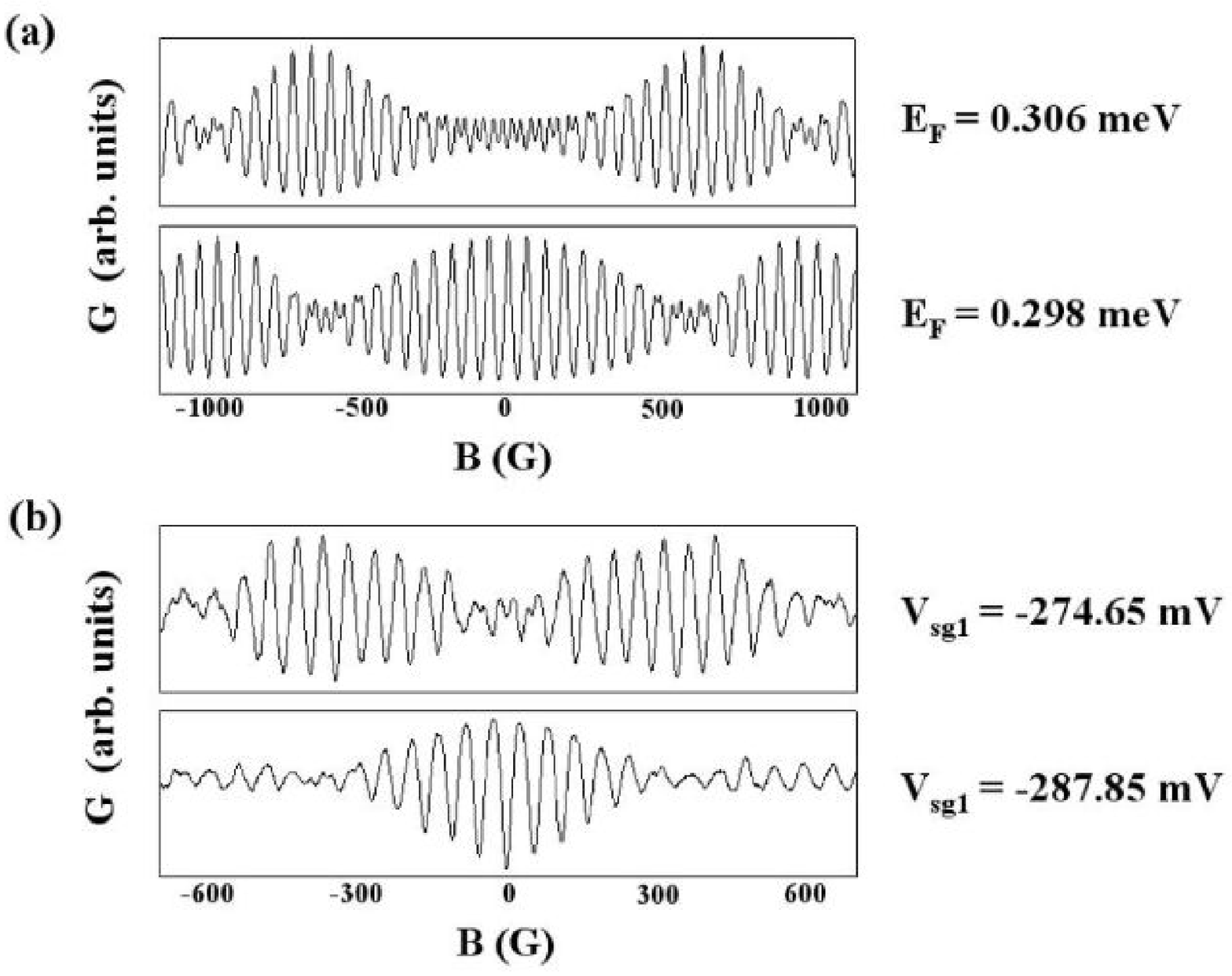}
\caption{}
\end{center}
\end{figure}

\begin{figure}[t]
\begin{center}
\leavevmode
\includegraphics[width=0.9\linewidth]{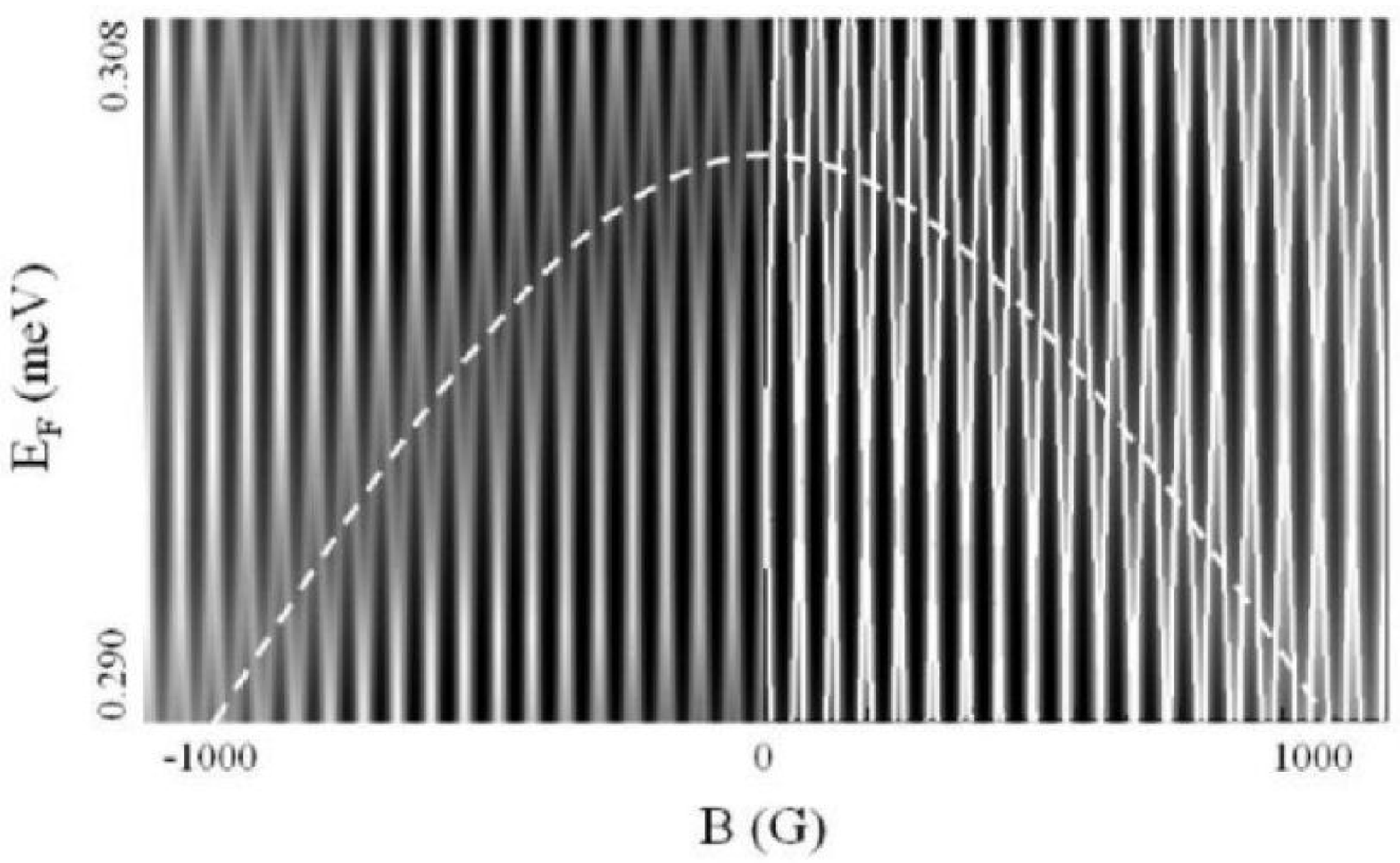}
\caption{}
\end{center}
\end{figure}

\begin{figure}[t]
\begin{center}
\leavevmode
\includegraphics[width=0.9\linewidth]{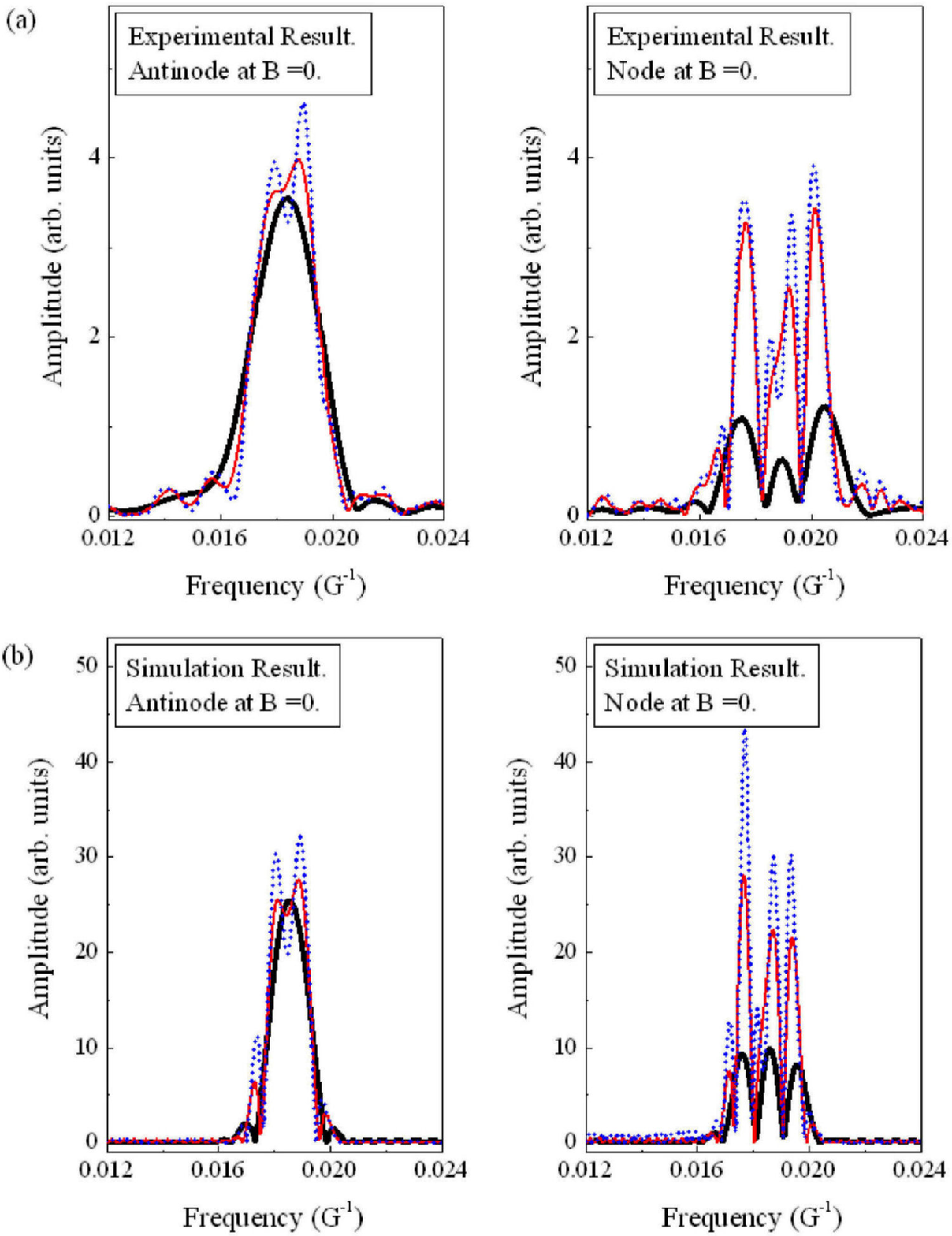}
\caption{}
\end{center}
\end{figure}


\begin{thebibliography}{00}
\bibitem{STSB} In our measurements, we first checked separately that the paths
between each of six gates (side gates 1,2,3,4, and quantum-dot
gates 1,2 in Fig. 1) and the island gate (in Fig. 1) gave a single
transverse sub-band mode by tuning each corresponding conductance
below $2e^2/h$. We then checked the conductance through the
interferometer as a whole to be lower than $2e^2/h$. This
confirmed us that the interferometer kept only a single transverse
sub-band mode. The resulting conductance with the subtraction of
the background revealed clear AB oscillations with beats even for
a low visibility.
\end{thebibliography}
\end{document}